\documentstyle [12pt] {article}
\begin {document}
\large
\begin {center}
{\bf About Possibility for Examination of Gravity Theories Using
the Precise Measurement of Particle Lifetime} \vspace{0.5cm}
\par
Kh.M. Beshtoev\\
\vspace{0.5cm} Joint Institute for Nuclear Research, Joliot Curie
6, 141980 Dubna,
Moscow region, Russia\\
\end {center}

\par
Abstract \\

\par
An approach for examination of gravitational theories using
precision measurements of particle lifetime is proposed. The
expressions describing dependence of particle lifetime on
gravitational potential in Einstein's and Newton's gravity
theories are obtained. In the case of Newton's gravity there is a
dependence of the particle velocity direction from the direction
of matter location, which creates the gravitational potential. If
the external gravitational field is spherical symmetric then there
would  be no possibility to distinguish these types of gravity. It
is found that the deposit of gravitational potential of the
Universe (in the case of uniformly distribution of matter in the
Universe) in particle lifetime is approximately one percent. On
the basis of the available experimental data it is found that
deposit of asymmetric gravitational field is $\frac{\varphi}{c^2}
\simeq 2 \cdot 10^{-4}$, i.e., if the experimental precision of
particle lifetime measurements will be several units of $10^{-4}$,
then we could see this effect.
\par
In reality, the lifetime of elementary particles  can be defined
by effective  masses of  these  particles in the external
gravitational field. The expressions for effective masses of
particles in the external gravitational field for two gravity type
theories are obtained. These masses can be used at computation of
the decay probability (or lifetime) of particles by standard
methods. It is shown that in this case it is also possible to
distinguish these two types of gravity theories\\

{\bf I. Introduction}\\

\par
Probably, brain of scientists has already settled the idea that
gravitational interactions are connected with space curvature and
our space is a curved one [1]. Nevertheless, this problem demands
a subsequent investigation.
\par
a). Gravitational Red Shift.
\par
In the work [2] it was shown, that radiation spectrum (or energy
levels) of atoms (or nuclei) in the gravitational field has a red
shift since the effective masses of radiant electrons (or
nucleons) change in this field. This red shift is equal to the red
shift of the radiation spectrum in the gravitational field
measured in existing experiments [3, 4]. The same shift must arise
when the photon (or $ \gamma $ quantum) is passing through the
gravitational field if it participates in gravitational
interactions. Then, on experiments one must detect double red
shift. Absence of  the double shift in the experiments means that
photons (or $ \gamma $ quanta) are passing through the
gravitational field without interactions (see also Ref. [5]).
\par
b). Advance of Perihelion of Planets.
\par
It is necessary to note that in the work [6, 7]  expression for
the advance of the perihelion of Mercury was obtained in the case
of flat space, which is mathematically equivalent to the
Einstein's expression.
\par
c). Photon Deflection in Gravitational Field.
\par
It is well known, that only massive bodies and particles
participate in the Newton's theory of gravitation (i.e. body and
particle having the rest mass).  Since the photons have no rest
mass, the usage of the "mass" $m_{ph}$ obtained in the formula
$$
m_ {ph} = \frac {E_ {ph}} {c^2} = \frac {h \nu} {c^2} ,
\eqno (1)
$$
is a hypothesis to be checked of.  The check has shown (see above
or [2] ) that there are no photons (or $ \gamma $ - quanta) red
shift when they pass through the gravitational field. It is
obvious, that since they have no rest masses (or a gravitational
charge), they cannot participate in the gravitational
interactions. Then, the deflection of photons in gravitational
fields must not exist either. The question arises: How could the
deflection of photons appear in the gravitational field, if they
do not participate in these interactions? It is clear that this
question calls for an answer. Let's note, that the given question
has been discussed in work [8] (see also references in [8]), where
it was shown that from the available experimental data it is not
impossible to come to a conclusion that photons are deflected in
the gravitational field.
\par
Probably, in order to clarify these problems it is necessary to
perform some experiments. I would like to indicate yet one more
experimental possibility for examination of the idea of connecting
gravitational interactions with space curvature (the Einstein's
theory of gravity).\\

{\bf II. About Possibility for Examination of Gravitational
Theories by Using Precise Lifetimes of Relativistic Particles}\\

\par
If particle's lifetime depends on gravity field (potential) in the
decay points, then we have a chance to examine how its lifetime
change in dependence of gravitational potential and velocity of
the particle. Further on, we will obtain an expression for
particle lifetime depending on gravitation potential and its
velocity in the framework of Einstein's and Newton's theory
gravity. It is necessary to stress that in this case we use the
gravitational potential in the point of particle location and it
includes the masses ($M$) of all objects, which create this
potential, i.e., in this approach will be used potential but
not potential gradient (see [3, 4]).\\

\par
{\bf II. 1. The Proper $\tau$  and World $x^o$ Times}\\

\par
In the General relativistic theory [1], it is supposed that
influence of gravitation is reduced to  appearance of a space-time
curvature described by the curvature tensor $g_{\alpha \beta};
\alpha, \beta = 0,1,2,3$, and the source of gravitational field is
the energy-momentum tensor $T_{\alpha \beta}$.  In case of a weak
gravitational field $\varphi/c^2 << 1$ from this curved space, we
may come to flat (quasi-flat) space, where gravitational field
contribution is determined by the potential $\varphi(x)$ [1].
Naturally, the $\varphi$ is equivalent to the Newton gravitational
potential. Particularly simple expressions are obtained in the
case of stationary (time independent) gravitational  field. In
this case the nondiagonal terms of curvature tensor are equal to
zero, and
$$
g_{o o} = 1 + 2\varphi/c^2  . \eqno(2)
$$
We will consider the case of stationary field since it presents
interest to us. Then connection between proper time $\tau$ (i.e.,
time when we take into account the gravity) and the world time
$x_o/c$ (the time without the gravity) is determined by the
following expression:
$$
x_o = \frac{\tau}{\sqrt{g_{oo}}} c = inv ,
\eqno(3)
$$
or
$$
\tau =\frac{1}{c} \sqrt{g_{oo}} x_o  .
\eqno(4)
$$
In the case of weak gravitational fields from (4), we come to the
following expression:
$$
\tau = \frac{x_o}{c} (1 + \frac{\varphi}{c^2}) . \eqno(5)
$$
\par
For the photon frequency $\omega$ in the external gravitational
field $\varphi$ we have the following expression
$$
\omega = \omega_o (1 - \frac{\varphi}{c^2}) , \eqno(6)
$$
where $\omega_o$ is photon frequency in the absence of
gravitational field.
\par
We will work in the framework of the method which is used in [1]
(L.D. Landau, E.M. Lifshitz) where $\varphi > 0$ and in the case
of necessity it needs to change on $\Delta \varphi$ which may be
either positive or negative.
\par
Since in the flat space the time $x_o$ duration is identical to
time $x_o'$ duration in the proper reference system of a physical
object, then the relativistic transformation between them has the
standard form
$$
x_o' = \gamma x_o  , \eqno(7)
$$
and the general formula for the proper time of relativistic object
in gravitational field takes the form
$$
x_o' = \gamma (1 + \frac{\varphi}{c^2}) x_o .
\eqno(8)
$$
Then for
$$
\tau' = \frac{x'_o}{c}, \qquad \tau_o = \frac{x_o}{c} ,
\eqno(9)
$$
we have
$$
\tau' = \gamma (1 + \frac{\varphi}{c^2}) \tau_o .
\eqno(10)
$$
The expression (10) describes duration of time for identical
physical processes of moving and a resting objects in the given
reference system in presence of gravity.
\par
We will consider $\tau'$ as a lifetime (or decay time) of the
particle with velocity $v$ in the stationary external
gravitational field $\varphi$.
\par
The photon frequency transformation, emitted by relativistic
object in gravitational field, is determined in a similar way
(however see Ref. [6, 7]).
\par
In the Newton's gravity case the analogous expression  for time
duration has the following form:
$$
\tau' = \gamma (1 + \frac{\varphi}{c^2} \frac{1}{\sqrt{1 + sin^2
\theta (\gamma^2 - 1)}}) \frac{x_o}{c}  , \eqno(11)
$$
where angle $\theta$ is angle between direction of the particle
velocity $\vec v$ and unity vector $\vec n$ to the center of
gravitational system.
\par
From Exp. (10) we can see that in Einstein's  case the
$\frac{\tau'}{ \gamma}$
$$
\frac{\tau'}{ \gamma} = (1 + \frac{\varphi}{c^2}) \frac{x_o}{c} =
inv  ,
\eqno(12)
$$
is  invariant. But in the Newton's case the value $\frac{\tau'}{
\gamma}$
$$
\frac{\tau'}{\gamma} = (1 + \frac{\varphi}{c^2} \frac{1}{\sqrt{1 +
sin^2 \theta (\gamma^2 - 1)}}) \frac{x_o}{c} , \eqno(13)
$$
is not invariant and depends on $\gamma$ and velocity direction.
\par
Now we have to discuss a question related to the required
precision for measurement of particles lifetime. For this purpose
we must know the estimation of value $\frac{\varphi}{c^2}$ created
by Universe matter in the experimental point (i.e. at the Earth
surface).
\par
Let us fulfill estimation of $\frac{\varphi}{c^2}$ of  our
Universe. For estimation of average value of $\varphi$ we can use
the following expression:
$$
d \varphi =  G \frac{dM}{R} =  G 4 \pi \rho \frac{R^2 dR}{R} =  4
\pi \rho R dR , \eqno(14)
$$
where $G$ is gravitational constant, $\rho$ is average matter
density of the Universe. Then
$$
\mid \bar \frac{\varphi}{c^2} \mid = 4 \pi G \rho \int^{R_o}_{0}
RdR = 2 \pi G \rho R^2_o ,
\eqno(15)
$$
where $\rho \sim 3.0 \cdot 10^{-31} g/cm^3$ is the Universe matter
density, $R_o$ is the Universe radius and $R_o \sim 10^{10}$ years
[9] (we presume that $c$ the light velocity equal to one). And
then the average value of $\frac{\varphi}{c^2}$ is
$$
\mid \bar \frac{\varphi}{c^2} \mid \simeq 1.25 \cdot 10^{-2} .
\eqno(16)
$$
Obviously, in the modern experiments we can principally reach one
percent precision and to measure the effects connected with the
gravitational field of the Universe matter. However, in previous
considerations we presumed that the matter is distributed
symmetrically in the Universe, and then we cannot distinguish the
Einstein's and Newton's gravity (see Exp. (11), (13)). In order to
distinguish, which of these gravity theories is realized in
Nature, the matter distribution in the Universe must be
asymmetrical. In  any case there is an asymmetry since the Earth
is not placed in the center of the Universe. At present, there are
found such asymmetries [9] and estimation of the gravitational
potential of this one gives that it is order of (if we take into
account the dark matter, i.e. neutrino masses and et cetera)
$$
\frac{\varphi}{c^2} \simeq  2 \cdot 10^{-4} . \eqno(17)
$$
From the Exp. (17), we can see that the experimental precision of
decay time measurements must be several units of $10^{-4}$  in
order to see
this effect.\\

{\bf II.2. Lifetime of Elementary Particles in a Stationary Gravitational Field }\\

\par
In reality, the lifetime of elementary particles  can be  defined
by effective  masses of  these  particles in external
gravitational field.  In this case we must use particle effective
masses, computing their lifetimes by the standard methods.
\par
Expression for energy of a physical object  in Einstein's theory
(when the object is moving along the world line) has the following
form [1]:
$$
E_o = m_o c^2 g_{oo} \frac{dx^o}{ds} = m_o c^2
\frac{dx^o}{\sqrt{g_{oo}} (dx^o)^2 - dl^2}. \eqno(18)
$$
Introducing the object velocity $v$ (i.e., observer time)
$$
v = \frac{dl}{d \tau} = \frac{c dl}{\sqrt{g_{oo}} dx^o} ,
\eqno(19)
$$
into the reference system of the observer, we obtain expression of
energy of the object in the given system
$$
E_o = \frac{m_o c^2 \sqrt{g_{oo}}}{\sqrt{1 - \frac{v^2}{c^2}}} .
\eqno(10)
$$
At $\frac{\varphi}{c^2} << 1$
$$
\sqrt g_{oo} \cong 1 + \frac{\varphi}{c^2} , \eqno(21)
$$
and we get
$$
E_o \cong  \frac{m c^2 (1 + \frac{\varphi}{c^2})}{\sqrt{1 -
\frac{v^2}{c^2}}} \qquad \Delta E =  \frac{m_o  \varphi}{\sqrt{1 -
\frac{v^2}{c^2}}}    . \eqno(22)
$$
From Exp. (22) we can come to a conclusion  that effective mass of
an object $M'$ (or a particle)  in external stationary
gravitational field is
$$
M' = M (1 + \frac{\varphi}{c^2})       . \eqno(23)
$$
\par
As we have already stressed above, in reality the lifetime of
elementary particles can be defined by effective  masses of  these
particles in the external gravitational field.  In this case, we
should use particle effective masses while  computing their
lifetimes. Naturally, this effect can be observed in  experiment.
\par
For example, lifetime $\tau (\pi)$ of $\pi^{\pm}$ mesons at their
lepton decays ($\pi \rightarrow l + \bar \nu_l$) is described by
the following expression:
$$
\tau (\pi) = \frac {1}{\Gamma (\pi)} ,
\eqno(24)
$$
where
$$
\Gamma (\pi) = \frac{G^2_F f_\pi^2 cos^2 \theta m^2_l m_\pi}{8
\pi} (1 - \frac{m^2_l}{m^2_\pi})^2 ,
\eqno(25)
$$
and $f_{\pi}$ is the pion decay constant, $G_F$ is Fermi constant,
$\theta$ is mixing angle, $m_l$  is lepton mass, $m_{\pi}$ is pion
mass.
\par
The lifetime of relativistic pion is defined by usage of standard
relativistic transformations. From Exp. (22), (23) it is  well
seen  that in the case of the Einstein's gravity, there must not
be any  dependency of the lifetime  of elementary  particle on the
external gravitational potential, i.e., there must not be a
visible effect.
\par
In the case of Newton's gravity, in contrast to the case of
Einstein's gravity, there must be dependence of external
gravitational potential as well as direction
and value of particle velocity, determined by Exp. (13).\\

{\bf III. Conclusion}\\

\par
This work proposes an approach for examination of gravitational
theories taking use of the precision lifetime measurements of
elementary particles. We obtained expression for dependence of
particle lifetimes of external gravitational potential in the case
of Einstein's and Newton's gravity. In the case of Newton's
gravity, there is a dependence of velocity direction from the
direction of matter location, which creates a gravitational
potential. If the external gravitational field is spherical
symmetric then there is no possibility to distinguish between
these types of gravity. It is found that the deposit of
gravitational potential of the Universe (in the case of uniformly
distribution of matter in the Universe) in particle lifetime is
about one percent. On basis of the available experimental data it
is found that deposit of asymmetric gravitational field is
$\frac{\varphi}{c^2} \simeq 2 \cdot 10^{-4}$, i.e., if the
experimental precision of particle lifetime measurements is
several units of $10^{-4}$, then we could see this effect and it
would be the confirmation of Newton's theory gravity (see
Introduction of this work).
\par
In reality, the lifetime of elementary particles  can be defined
by effective  masses of  these  particles in the external
gravitational field. The expressions for effective masses of
particles in the external gravitational field for two gravity type
theories are obtained. These masses can be used at computation of
the decay probability (or lifetime) of particles by standard
methods. It is shown that in this case it is also possible to
distinguish these two types of gravity theories.
\par
The author expresses his gratitude to A. Sapogov, a co-author of
the
work [10], for discussions.\\

\par
{\bf References}\\

\par
\noindent 1. A. Einstein, Ann. Phys. (Leipzig) {\bf 49}, 769,
(1916);
\par
 L.D. Landau, E.M. Lifshitz, Field Theory, M., Nauka, 1988, p.324.
\par
\noindent 2. Kh. Beshtoev, Physics Essays {\bf 13}, N3, (2000).
\par
\noindent 3.  R.V. Pound, G.A. Rebka, Phys. Rev. Let. {\bf 4},
337, (1960);
\par
 R.V. Pound,  J.L. Snider, Phys. Rev. {\bf 140}, 788, (1965).
\par
\noindent 4.  J.L. Snider, Phys. Rev. Let. {\bf 28}, 853, (1972).
\par
\noindent 5. P. Marmet, Einstein's Theory of Relativity versus
Classic
\par
Mechanics, Newton Physics Books, Canada, 1997.
\par
\noindent 6. P. Marmet, Physics Essays {\bf 12}, N3, (1999).
\par
\noindent 7. Kh. Beshtoev, JINR Communication E2-2001-107, Dubna,
2001.
\par
\noindent 8. P. Marmet and C. Couture, Physics Essays {\bf12},
162,(1999).
\par
\noindent 9. W. C. Saslaw, Gravitational Physics of Stellar and
Galactic Systems, Cambridge, Cambridge Univ. Press, 1987.
\par
A. Dressler, Scientific American N11, 1987.
\par
\noindent 10.  Kh Beshtoev et all., JINR Communication E2-2003-18,
2003, Dubna.

\end{document}